\documentclass[a4paper]{jpconf}
\usepackage{graphicx}
\def\nub        {\overline{\nu}}

\def\nue        {\nu_e}

\def\num        {\nu_\mu}

\begin{document}
\title{The WAGASCI detector as an off-axis near detector of the T2K and Hyper-Kamiokande experiments}

\author{Benjamin Quilain, Akihiro Minamino for the WAGASCI, T2K and HK collaborations}

\address{Production Editor, \jpcs, \iopp, Dirac House, Temple Back, Bristol BS1~6BE, UK}

\ead{bquilain@scphys.kyoto-u.ac.jp}

\begin{abstract}
In the search for CP violation at the T2K and future Hyper-Kamiokande experiments, it is crucial to reduce the present systematic uncertainties. The current T2K near detector, ND280, reduces the uncertainties coming from the neutrino beam and cross-section models from $11.9\%$ to $5.4\%$ in the $\nu_{e}$ appearance channel. These residual uncertainties mostly come from intrinsic limitations of ND280 due to its difference in target material and angular acceptance with the far detector.\\
In order to show evidence (subsequently observation) of the CP violation in the T2K phase-II and Hyper-Kamiokande experiments, this paper proposes an upgrade of the ND280 detector. It uses a 3D grid scintillator structure surrounded by Time Projecting Chambers in order to reconstruct particles with $\sim 4\pi$ acceptance.
\end{abstract}
\section{Motivations for a near detector upgrade}
\begin{figure}[h]
\begin{minipage}{20pc}
\includegraphics[width=20pc]{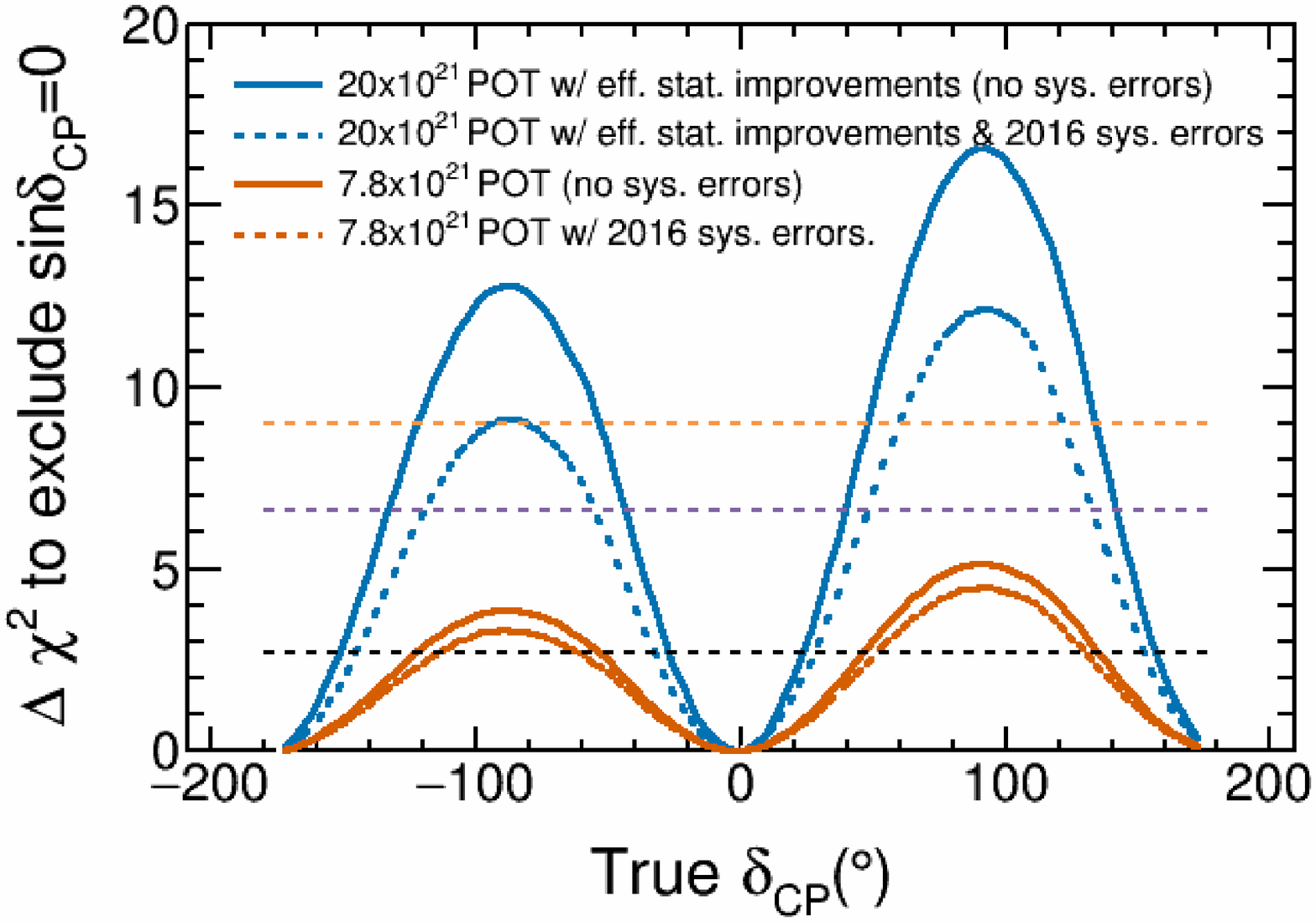}
\caption{\label{CPSensitivityT2K} Exclusion contour of CP conservation as a function of the true $\delta_{CP}$ value if the mass hierarchy is known (true Normal Hierachy).}
\end{minipage}\hspace{1pc}%
\begin{minipage}{20pc}
\includegraphics[width=20pc]{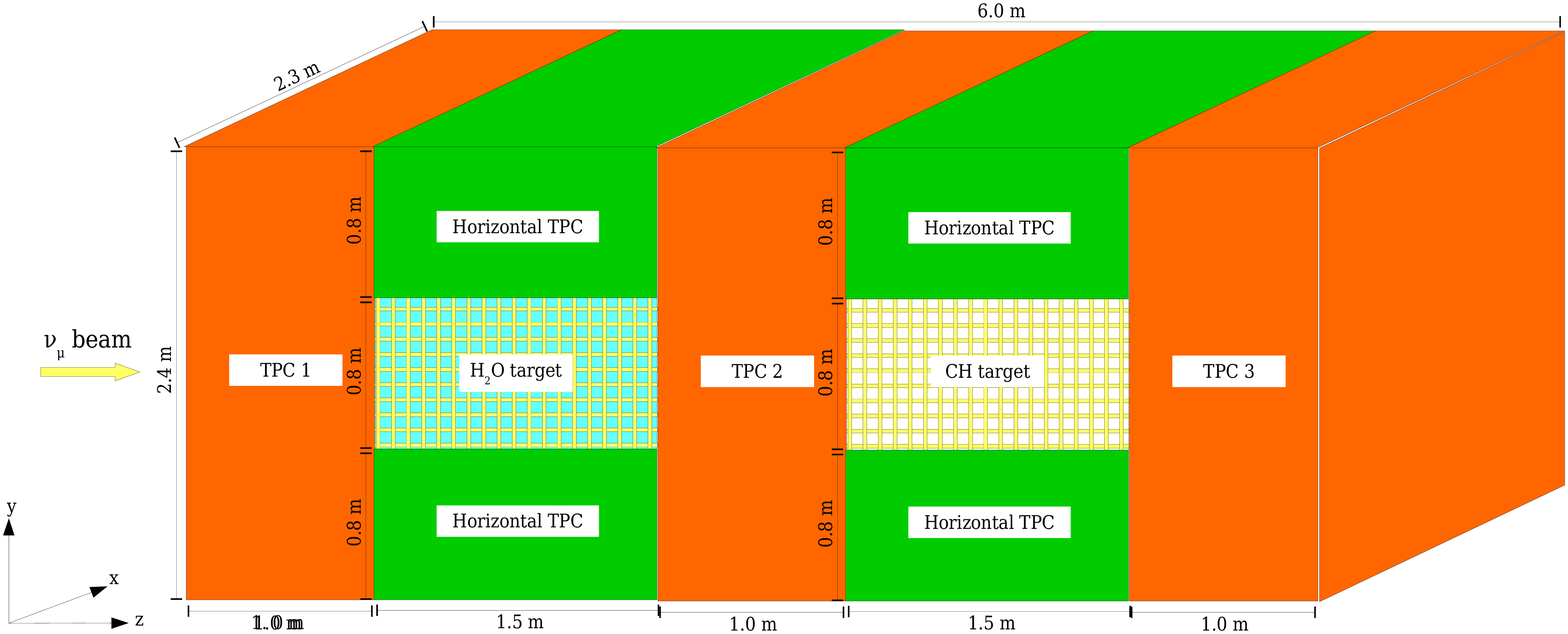}
\caption{\label{ND280Design} Schematic view of the ND280-upgrade detector.}
\end{minipage} 
\end{figure}

T2K (Tokai-to-Kamioka)~\cite{abe2011t2k,abe2014observation} is a long-baseline off-axis neutrino oscillation experiment that focuses on measuring $\num \rightarrow \num$ and $\num \rightarrow \nue$ in both $\nu$ and $\bar{\nu}$-modes. A very pure $\num$ beam is produced at J-PARC accelerator complex and detected 295~km away at the Super-Kamiokande (SK)~\cite{fukuda2003super} far detector. T2K uses a set of near detectors in order to reduce the large uncertainties on the oscillation parameters that come from the neutrino fluxes and interaction models. The ND280 off-axis detector~\cite{abe2011t2k} measures the non-oscillated $\num$ spectrum, constrains the cross-section models and determines the $\nue$ intrinsic contamination in the beam. It reduced the uncertainties from $12.0\%$ to $5.0\%$ and from $11.9\%$ to $5.4\%$ respectively in the $\num \rightarrow \num$ and $\num \rightarrow \nue$ channels. However, this systematic uncertainty reduction is limited due to intrinsic detector limitations:
\begin{itemize}
\item differences in most of the target material between ND280 (80$\%$ C$_{8}$H$_{8}$+20$\%$ H$_{2}$O) and SK (pure H$_{2}$O). This leads to using the cross-section model to cover the difference between C$_{8}$H$_{8}$ and H$_{2}$O.
\item differences in the angular acceptance of ND280 (mostly foward) and SK ($4\pi$). The flux and cross-section models are therefore not constrained in the high angle regions.
\item inability to reconstruct low momentum protons ($<450$ MeV/c) at the near detector. This leads to a large uncertainty on the interaction of the neutrino with bounded states of 2 protons (2p-2h) and Final State Interactions (FSI) that affect the far detector differently than ND280 due to the different threshold in proton momentum.
\item differences in the flux at the near and far detector.
\end{itemize}
Figure~\ref{CPSensitivityT2K} shows how the current systematic error limits the phase space corresponding to a $3\sigma$ evidence of CP violation at the T2K phase II experiment~\cite{abe2016proposal}. In order to remove the impact of possible statistical fluctuations, mass hierarchy or true value of $\delta_{CP}$, we propose to upgrade the ND280 near detectors that will tackle the previously listed issues while keeping the current advantages of the detector. Note that the flux difference will remain but its fractional error is lower than $1\%$ in the oscillation maximum. This upgrade will also be crucial for observing CP violation at the future Hyper-Kamiokande experiment with more than $5\sigma$ confidence level.

\section{The ND280-upgrade detector}
The ND280-upgrade design is shown in Figure~\ref{ND280Design}. It is constructed based on a target embedded in a central tracker that is surrounded by Time Projecting Chambers (TPCs). 
\begin{figure}[h]
\begin{minipage}{25pc}
\includegraphics[width=25pc]{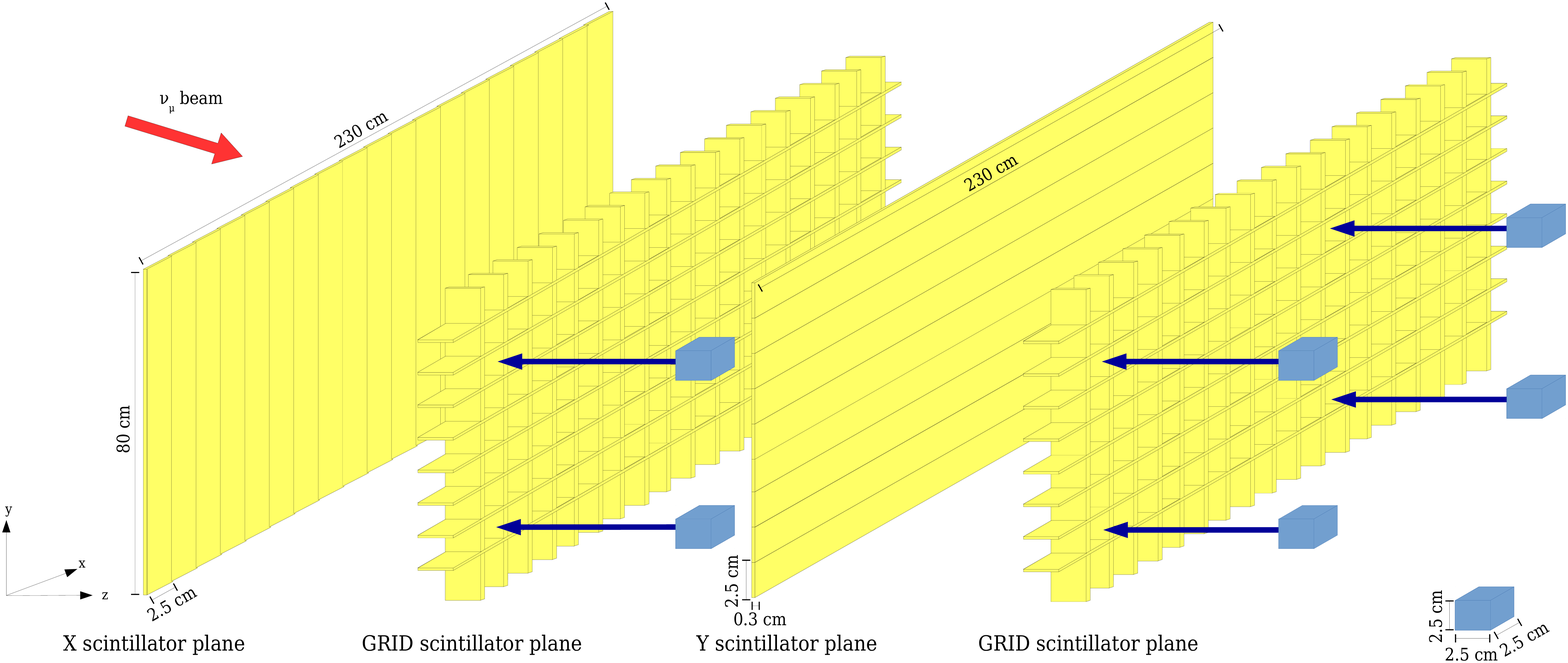}
\caption{\label{3DGrid} The ND280-upgrade tracker.}
\end{minipage}\hspace{1pc}%
\begin{minipage}{17pc}
\includegraphics[width=17pc]{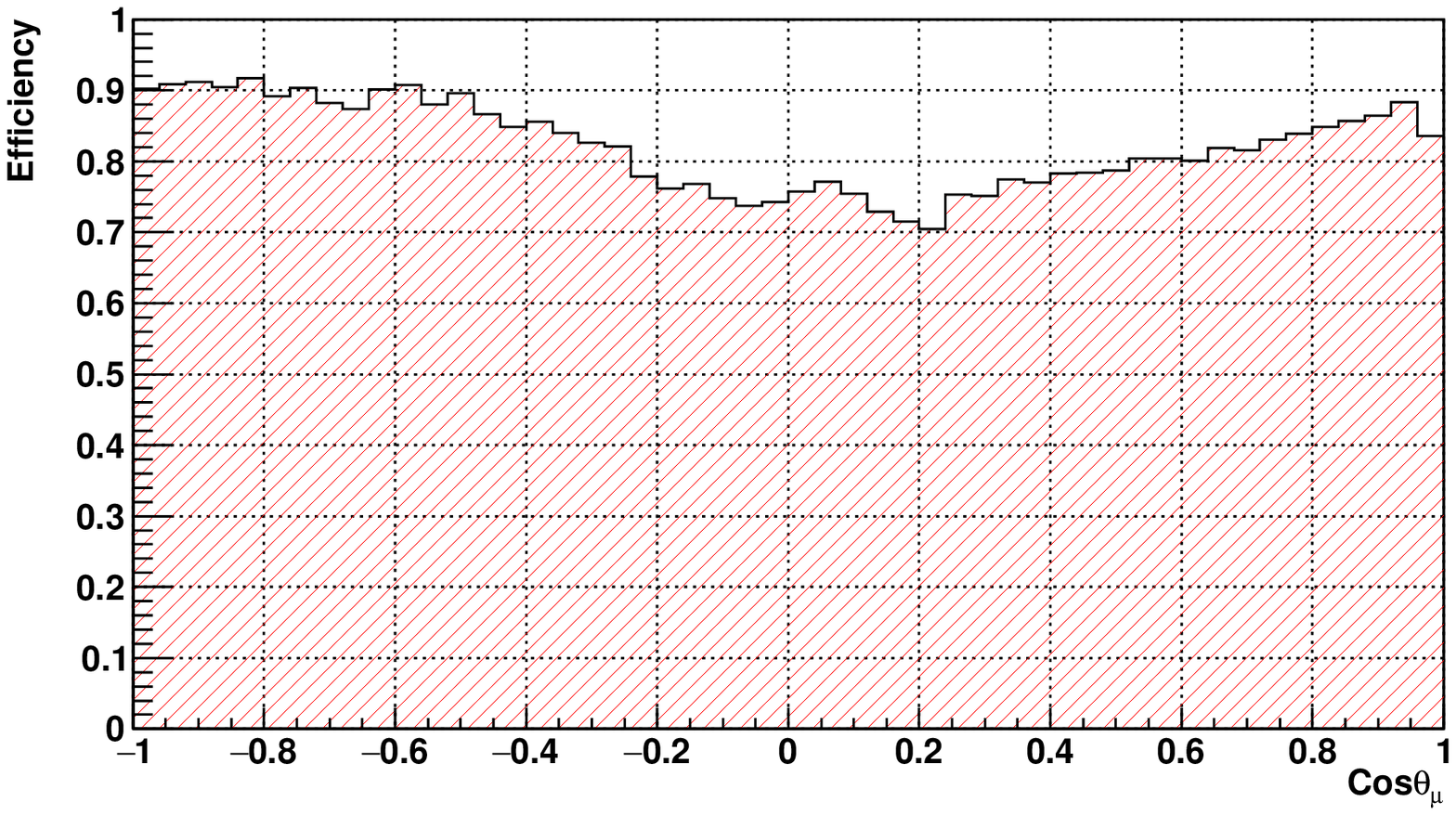}
\caption{\label{EfficiencyAngle} Reconstruction efficiency as a function of muon angle.}
\end{minipage} 
\end{figure}
Two different modules are considered:
\begin{itemize}
\item A 2.8~tonne H$_{2}$O module to mimic the target of the far detector. A scintillator tracker (see description below) is embedded in this module. In the current configuration, the module is comprised of 70$\%$ H$_{2}$O and 30$\%$ C$_{8}$H$_{8}$.
\item A 0.9~tonne C$_{8}$H$_{8}$ module consisting only of scintillators. It allows the scintillator background in the first module to be substracted. Moreover, a fully active detector is crucial in order to reconstruct low momentum protons.
\end{itemize}
The central tracker is designed in order to have an excellent reconstruction of large angle tracks to mimic the far detector 4$\pi$ acceptance. It is constructed by alternating vertically and horizontally aligned scintillator planes for tracking forward going tracks (similar to the current ND280), with new 3D GRID scintillator planes which allow the reconstruction of high angle tracks (see Figure~\ref{3DGrid}).\\
The TPCs are responsible for separating $\nu$ and $\nub$ components, identifying the interaction topology and measuring the neutrino energy. The latter is measured through the identification of the daughter particles via measurement of their charge and kinematics. The TPCs are only located upstream and downstream of the tracker in the current ND280 which limits the high angle acceptance. It is therefore proposed to add horizontal TPCs on the top and bottom of the tracker in order to extend their acceptance to $4\pi$ for the upgraded detector. The size of these additional TPCs is tuned coarsely to 80~cm in order to adapt the current TPCs characteristics for the high angle muons that have lower momenta. 


\section{Water-in and water-out detector performance}
The performance of the upgrade detector is established using a reconstruction adapted from the WAGASCI detector~\cite{koga2015water}. The water cell size is first optimized to cubic cells of 2.5~cm side in order to keep a high water content while reducing the threshold in particle momenta. Though the detector is not fully active, it ensures that over $70\%$ of the muons above $p_{\mu}>50$~MeV/c are reconstructed. Note that the oscillation maximum concerns neutrinos having $E_{\nu} \sim 600~$MeV. The new 3D grid ensures a very high reconstruction efficiency ($>70\%$) even for large angle muons as shown on Figure~\ref{EfficiencyAngle}. The ND280-upgrade is therefore an ideal detector to cover the SK phase space.
Note that since the current ND280 topology is not only based on muon but also on pion identification, it is crucial to ensure an excellent pion reconstruction and identification. This criterion is currently used for final optimization of the detector size and granularity.
\subsection*{The water-out detector}
A C$_{8}$H$_{8}$ detector is needed in order to substract the scintillator background in the water-in module. Moreover, the large contamination of 2p-2h interactions in the CCQE sample leads to a $>2\%$ uncertainty in the $\nue$ appearance channel. The upgrade aims to reduce this uncertainty through separation of these two topologies, using proton counting and kinematics. However, the water-in detector is unable to reconstruct low momentum protons ($p_{p} \geq 450~$MeV/c) which constitutes $\sim 50\%$ of the sample. Therefore, we propose a water-out detector (see Figure~\ref{ND280Design}). Using the current reconstruction, it lowers the threshold on proton momentum to 200-250~MeV/c. Consequently, $>80\%$ of the protons produced in CCQE and 2p-2h interactions are reconstructed. On top of pure proton counting, it opens very promizing possibilities for separating the CCQE and 2p-2h interactions by using single and double transverse variables~\cite{lu2015measurement,lu2015reconstruction}.


\section*{References}
\begin{thebibliography}{9}
\bibitem{abe2011t2k}
T2K collaboration, K.\ Abe \etal, Nucl.\ Instr.\ Meth.\ Phys.\ Rev.\ A {\bf 659(1)}, 106--135 (2011).

\bibitem{abe2014observation}
T2K collaboration,
K.\ Abe \etal, Phys.\ Rev.\ Lett.\ {\bf 112(6)}, 061802 (2014).

\bibitem{fukuda2003super}
Super-Kamiokande collaboration,
S.\ Fukuda \etal, Nucl.\ Instr.\ Meth.\ Phys.\ Rev.\ A {\bf 501(12)}, 418--462 (2003).

\bibitem{fukuda2003super}
Super-Kamiokande collaboration,
S.\ Fukuda \etal, Nucl.\ Instr.\ Meth.\ Phys.\ Rev.\ A {\bf 501(12)}, 418--462 (2003).

\bibitem{abe2016proposal}
T2K collaboration,
K.\ Abe \etal, Proposal for an Extended Run of T2K to $20 \times10^{21}$ POT{\it Preprint} hep-ex/1609.04111.

\bibitem{koga2015water}
T. Koga \etal, Proceedings of the 2nd International Symposium on Science at J-PARC---Unlocking the Mysteries of Life, Matter and the Universe---, id. 023003, 6 pp., 1 (2015).

\bibitem{lu2015measurement}
X-G. Lu, Measurement of nuclear effects in neutrino interactions with minimal dependence on neutrino energy, {\it Preprint} hep-ex/1512.05748.

\bibitem{lu2015reconstruction}
X-G. Lu \etal,  , Phys.\ Rev.\ D\ {\bf 92(5)}, 051302 (2015).

\end{thebibliography}
\end{document}